\documentclass[universe,article,accept,moreauthors,pdftex,10pt,a4paper]{mdpi} 
\firstpage{1} 
\articlenumber{x}
\doinum{10.3390/------}
\pubvolume{xx}
\pubyear{2017}
\copyrightyear{2017}
\externaleditor{}
\history{Received: October 31, 2017}

\usepackage{amssymb}

\Title{Pure-connection gravity and anisotropic singularities}

\Author{Kirill Krasnov $^{1}$\orcidA{} and Yuri Shtanov $^{2}$\orcidB{}}

\AuthorNames{Kirill Krasnov and Yuri Shtanov}

\address{%
	$^{1}$ \quad School of Mathematical Sciences, University of Nottingham,
	University Park, Nottingham, NG7 2RD, UK; Kirill.Krasnov@nottingham.ac.uk\\
	$^{2}$ \quad Bogolyubov Institute for Theoretical Physics, Metrologichna St.\@ 14-b, Kiev 03680, Ukraine; shtanov@bitp.kiev.ua}

\abstract{In four space-time dimensions, there exists a special infinite-parameter family of chiral
modified gravity theories. They are most properly described by a connection field, with space-time 
metric being a secondary and derived concept. All these theories have the same number of degrees 
of freedom as general relativity, which is the only parity-invariant member of this family.  Modifications 
of general relativity can be arranged so as to become important in regions with large curvature.  In this 
paper we review how a certain simple modification of this sort can resolve the Schwarzschild black-hole 
and Kasner anisotropic singularities of general relativity.  In the corresponding solutions, the fundamental 
connection field is regular in space-time.}


\begin{document}


\section{Introduction}

Modification of general relativity theory has long been the subject of many investigations.  
One of the simplest examples is the $R^2$ gravity, of relevance as a valid model of inflation
\cite{Starobinsky:1980te, Ade:2015lrj}. This is a classical example of the scalar-tensor theory, 
propagating not just the two polarizations of the graviton but also a scalar. More involved 
modifications of GR with higher powers of the curvature added to the Lagrangian
propagate more degrees of freedom. It seems impossible to modify metric-based GR 
without adding extra propagating DOF\@. This is the content of several GR uniqueness 
theorems available in the literature.

The situation changes dramatically when, instead of metric, one adopts different variables for 
describing gravitational degrees of freedom.  Thus, it turns out possible to modify GR without adding
new degrees of freedom if one starts from one of its chiral descriptions based on spin connection, 
with metric becoming a secondary and derived object.  This 
formulation of GR originated from the seminal work due to Pleba\'nski \cite{Plebanski:1977zz}, 
but its pure-connection form
and its modifications with the mentioned properties are, perhaps, less known.   In fact, it gives birth to
an {\em infinite-parametric\/} class of chiral modified gravity theories without new DOF, in which
GR is just a special member \cite{Bengtsson:1990qg, Krasnov:2006du}.  

When studying these modified-gravity theories in some particular setups, we came across their interesting
features in relation to certain anisotropic singularities encountered in general relativity.  Specifically, 
the black-hole singularity and the Kasner singularity of Bianchi~I space-time can, in a sense, be 
`resolved' in a class of modified theories under investigation.  Although
the metric field based on this solution still contains singularities and
experiences changes of signature, the fundamental connection field is everywhere regular.
In this talk, we will review this feature of the theory under investigation.

\section{Pure-connection gravity and its modification}


\subsection{Eddington--Schr\"{o}dinger theory}

This is historically the first pure-connection formulation of GR\@. 
A simple way to get it is to start from the first-order Palatini formulation,
with the affine connection $\Gamma_{\mu\nu}{}^\rho$ and the metric $g_{\mu\nu}$ being independent variables. 
Integrating the variables $\Gamma_{\mu\nu}{}^\rho$, one obtains the Hilbert--Einstein action for the metric.
On the other hand, `integrating out' the metric $g_{\mu\nu}$, i.e., solving the field equations for $g_{\mu\nu}$ and
substituting the solution back into the action, one obtains a pure-connection theory that contains only 
$\Gamma_{\mu\nu}{}^\rho$. 
This procedure is only possible in the presence of a non-zero cosmological constant. In the Eddington--Schr\"{o}dinger
case \cite{Eddington, Schrodinger}, the pure-connection action is
\begin{equation}\label{Eddington}
S[\Gamma] = \frac{1}{8\pi G\Lambda} \int \sqrt{ \det  \left( R_{\mu\nu}[\Gamma]
	\right)}\, d^4 x \, ,
\end{equation}
where the symmetric Ricci tensor $R_{\mu\nu}$ is constructed from $\Gamma_{\mu\nu}{}^\rho$ only.
The field equations that result from (\ref{Eddington}) are
\begin{equation}\label{E-feqs}
\nabla_\rho R^{\mu\nu}[\Gamma] = 0\, ,
\end{equation}
where $R^{\mu\nu}[\Gamma]$ is the inverse of $R_{\mu\nu}[\Gamma]$. If one now makes a definition
\begin{equation} \label{g-gam}
g_{\mu\nu} = \frac{1}{\Lambda} R_{\mu\nu}[\Gamma]\, ,
\end{equation}
then equation (\ref{E-feqs}) tells that $\Gamma_{\mu\nu}{}^\rho$ is the
$g$-compatible connection. The definition (\ref{g-gam}) of $g_{\mu\nu}$ is then the
vacuum Einstein equation.

\subsection{The chiral Pleba\'nski formulation of GR}

In the chiral approach, one starts with the Pleba\'nski formulation of GR
\cite{Plebanski:1977zz} (see also \cite{Capovilla:1991qb, Krasnov:2009pu}). This is a
first-order formulation, with an $\mathfrak{so} (3)$ Lie-algebra-valued two-form field
and a connection as independent variables. This formulation is going to be the basis for all
constructions below.

The Lie-algebra indices are denoted by lower-case Latin letters $i,j,k,\ldots =1,2,3$. The
basic fields are a Lie-algebra-valued two-form field with components $B^i$ and a
connection one-form with components $A^i$. There is also a Lagrange multiplier field
$\Psi^{ij}$, which is symmetric and traceless. The action of the theory is
\begin{equation}\label{plebanski}
S[B,A,\Psi] = {\rm i} \int B^i \wedge F^i - \frac{1}{2} \left(\Psi^{ij} +
\frac{\Lambda}{3}\delta^{ij}\right) B^i \wedge B^j \, .
\end{equation}
Here, $\Lambda$ is the cosmological constant, which may be zero. The imaginary unit
in front of the action is needed in order to make it real for fields
satisfying the reality conditions as appropriate for Lorentzian signature; see below. 

Let us consider the field equations stemming from (\ref{plebanski}). First, varying with
respect to $B^i$, we get
\begin{equation}\label{Pleb-main-eqn}
F^i = \left(\Psi^{ij} + \frac{\Lambda}{3}\delta^{ij}\right) B^j \, .
\end{equation}
The Euler--Lagrange equation for the connection is
\begin{equation}\label{compat}
d_A B^i = 0 \, ,
\end{equation}
where $d_A$ is the covariant exterior derivative with respect to the connection $A^i$. 
Finally, there is an equation obtained by varying with respect to $\Psi^{ij}$.  Since the matrix-valued 
	field $\Psi^{ij}$ is constrained to be symmetric and traceless, variation of (\ref{plebanski}) 
	with respect to this field tells that the traceless part of the matrix-valued four-form 
	$B^i \wedge B^j$ vanishes.  In other words, the matrix $B^i \wedge B^j$ is proportional to
	the identity matrix $\delta^{ij}$, so that we obtain
	\begin{equation}\label{simpl}
		B^i \wedge B^j = \frac13 \delta^{ij} B^k \wedge B^k \, .
\end{equation}
This equation can be understood as telling that $B^i$ `come from a tetrad' in the sense
that $B^i$ satisfying this equation contain no more information than that provided by the
metric plus a choice of an ${\rm SO}(3)$ frame at every space-time point.  This metric is defined 
by the two-form fields via the Urbantke formula \cite{Urbantke:1984eb}:
\begin{equation}\label{urb}
g_{\mu\nu} \sqrt{\det  g} \propto \tilde{\epsilon}^{\alpha\beta\gamma\delta} \epsilon^{ijk}
B^i_{\mu\alpha} B^j_{\nu\beta} B^k_{\gamma\delta}\, .
\end{equation}
Here $\tilde{\epsilon}^{\alpha\beta\gamma\delta}$ is the anti-symmetric tensor density of
weight one which in any coordinate system has components $\pm 1$, and the proportionality
means equality up to an arbitrary positive coordinate-dependent scalar factor. To fix the
metric completely, it suffices to fix this factor, or to specify the associated volume
form. In the case of GR, in view of relation (\ref{simpl}), the associated volume form
${\mathcal V}_g$ is unique up to a constant, which is fixed by the requirement 
that the Einstein equations are satisfied with the cosmological constant $\Lambda$:
\begin{equation}\label{vol-BB}
3!\, {\rm i}\, {\mathcal V}_g = B^i\wedge B^i \equiv {\rm Tr\,} B \wedge B \, .
\end{equation}

Equation
(\ref{compat}) can be solved for $A^i$ in terms of derivatives of $B^i$ whenever $B^i$
are non-degenerate (that is, the three two-forms $B^i$ are linearly independent). In
particular, equation (\ref{compat}) can be solved if the two-forms $B^i$ satisfy
(\ref{simpl}), in which case the solution $A^i$ can be shown to be just the self-dual
part of the Levi-Civita connection for the metric described by $B^i$. Equation
(\ref{Pleb-main-eqn}) then becomes a statement that the curvature of the self-dual part
of the Levi-Civita connection of a metric is self-dual as a two-form.
This is equivalent to the Einstein condition, which shows that (\ref{plebanski}) is indeed a description of
GR\@. When all field equations are satisfied, the field $\Psi^{ij}$ is
identified with the self-dual part of the Weyl curvature. For more details on this
formulation, the reader is referred to \cite{Krasnov:2009pu}.

\subsection{Modified gravity}

A significant feature of the Pleba\'nski formulation (\ref{plebanski}) of general relativity is that 
it allows for modifications (or deformations) that do not increase the number of its degrees of freedom 
\cite{Krasnov:2006du}. Specifically, it consists in allowing the cosmological constant in (\ref{plebanski}) 
to be an arbitrary ${\rm SO} (3)$-invariant function of the field $\Psi^{ij}$\,:
\begin{equation}\label{pleb-modified}
S[B,A,\Psi] = {\rm i} \int B^i \wedge F^i - \frac{1}{2} \left(\Psi^{ij} +
\frac{\Lambda(\Psi)}{3}\delta^{ij}\right) B^i \wedge B^j \, .
\end{equation}
It can be shown \cite{Krasnov:2008zz} by the Hamiltonian analysis that this theory
continues to propagate just two degrees of freedom, similarly to GR\@. At the same time,
this is a modified theory of gravity, in which modification becomes important in
space-time regions where the function $\Lambda(\Psi)$ significantly deviates from a
constant. A particularly simple one-parameter family of modifications is obtained by
considering the function $\Lambda(\Psi)$ in the form of a quadratic polynomial in
$\Psi^{ij}$\,:
\begin{equation}\label{alpha-family}
\Lambda(\Psi) = \Lambda_0 - \frac{\alpha}{2} {\rm Tr\,} (\Psi^2) \, ,
\end{equation}
where $\alpha > 0$ is an arbitrary parameter.
For this family of modified theories, one expects strong deviations from GR when the Weyl
curvature $\Psi$ becomes of the order of $\sqrt{\Lambda_0 /\alpha}$.

\subsection{Pure-connection formulation}

The pure-connection formulation can be obtained from (\ref{pleb-modified}) by integrating out
all variables except connection.  This is possible since the auxiliary fields enter without derivatives
in the action.  Eventually, one obtains action in the form
\begin{equation}\label{S-pure-conn}
S[A] = \frac{{\rm i}}{2} \int {\mathcal L} (X) {\mathcal V} \, ,
\end{equation}
where ${\mathcal V}$ is an arbitrary volume form, the matrix field $X^{ij}$ is defined by
\begin{equation}\label{X}
F^i \wedge F^j = X^{ij} {\mathcal V} \, ,
\end{equation}
and ${\mathcal L} (X)$ is a homogeneous ${\rm SO} (3)$-invariant function of $X$. 
One can see that (\ref{S-pure-conn}) is, in fact, independent of the volume form ${\mathcal V}$.
The homogeneous function ${\mathcal L} (X)$ is in one-to one correspondence with the function 
$\Lambda (\Psi)$ in (\ref{pleb-modified}), although this correspondence is not easy to find explicitly.
General relativity [with $\Lambda (\Psi) \equiv \Lambda_0$] corresponds to 
\cite{Krasnov:2011pp}
\begin{equation} \label{GR}
{\mathcal L}_{\rm GR} (X) = \frac{1}{\Lambda_0} \left( {\rm Tr\,} \sqrt{X} \right)^2 \, .
\end{equation}

The pure-connection equation of motion for the theory under consideration is the second-order
partial differential equation
\begin{equation}\label{eqn-pure-conn}
d_A \left( \frac{\partial {\mathcal L}}{\partial X^{ij}} F^j \right) = 0 \, .
\end{equation}

The volume form of the metric arising in general relativity is specified by (\ref{vol-BB}).  
It is proportional to the Lagrangian volume form ${\mathcal L} \left( F \wedge F \right) \equiv 
{\mathcal L} (X) {\mathcal V}$.
For a modified gravity theory with a general function ${\mathcal L} ( X )$ as a
Lagrangian, it is not clear which metric from the conformal class determined by (\ref{urb}) should
be interpreted as the `physical' one. By analogy with GR, we could assume that its volume form 
will remain to be proportional to ${\rm Tr\,} B \wedge B$, 
or that it will be proportional to the action density ${\mathcal L} \left( F\wedge F \right)$.  
These two options no longer coincide in a modified theory of gravity and can be regarded as 
equally plausible.

\section{Black-hole solution}

In \cite{Krasnov:2007ky, Krasnov:2008sb}, we obtained and described a complete vacuum solution 
in theory (\ref{pleb-modified}) with spherical symmetry.  It turns out that the solution respects the Birkhoff 
theorem: it is necessarily static.  This is another manifestation of the absence of new degrees of freedom in
the theory.  Due to spherical symmetry, the symmetric traceless field $\Psi^{ij}$ has the form
\begin{equation}
\Psi^{ij} = \psi (r) \left( 3 \frac{x^i x^j}{r^2} - \delta^{ij} \right)	\, ,
\end{equation}
and is described by a single function $\psi (r)$ of the radial coordinate $r = \sqrt{\sum_i (x^i)^2}$.  
The cosmological function $\Lambda (\Psi)$ becomes a function of $\psi$, and its derivative with 
respect to $\psi$, which we denote by $\Lambda_\psi$, quantifies the deviation of the theory from GR:
The condition $\left| \Lambda_\psi \right| \ll 1$  implies the validity of GR\@.  

The solution for the gauge field can be found in \cite{Krasnov:2007ky} and is expressed as
\begin{equation}
A^1 = f (r) \sin \theta d \phi \, , \qquad A^2 = - f (r) d \theta \, , \qquad A^3 = {\rm i} \frac{r r_g (\infty)}{r_g (r)} 
\left[ \psi (r) - \frac{\Lambda(r)}{3}  \right] d t - \cos \theta d \phi \, ,
\end{equation}
where $\Lambda (r) \equiv \Lambda \left( \psi (r) \right)$,
\begin{equation}
f (r) = \sqrt{1 - \frac{r_g (r)}{r} - \frac13 \Lambda (r) r^2} \, , \qquad
\psi (r) = \frac{r_g (r)}{2 r^3} \, ,
\end{equation}
and the function $r_g (r)$ is determined from the equation
\begin{equation}
\quad r_g (r) \exp \left( \int_0^{r_g (r)/2 r^3}
\frac{\Lambda_\psi}{6 \psi} d \psi \right) =r_g (\infty)	\, .
\end{equation}
The arising metric, specified by the condition of anti-self-duality
of the curvature components $F^i$ and by the metric volume form ${\mathcal V}_g \propto {\rm Tr\,} B \wedge B$,
has the following form:
\begin{equation}\label{metric}
ds^2 = \left( \frac{1 - \Lambda_\psi / 3}{1 + \Lambda_\psi / 6}\right) \left[
\frac{r_g^2 (\infty)}{r_g^2 (r)} f^2 (r) d t^2 - f^{-2} (r) d r^2 \right] - r^2 d \Omega^2 \, .
\end{equation}

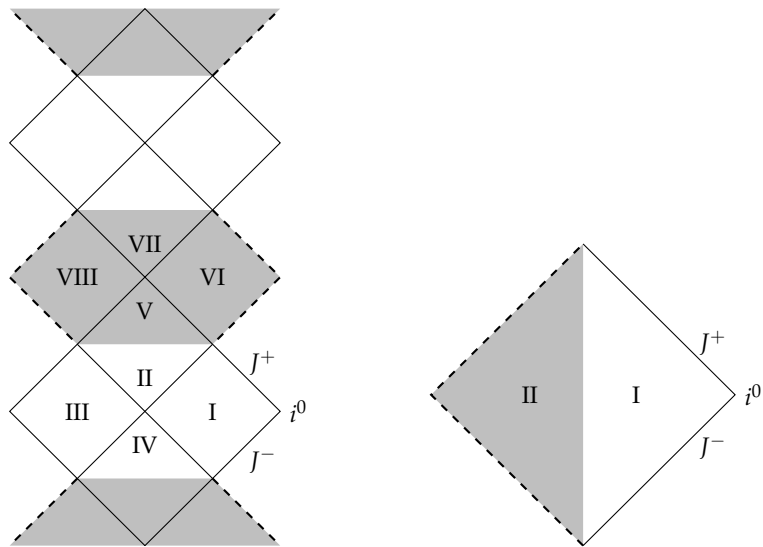
\begin{figure}[H]
	\centering
\begin{tikzpicture}[scale=0.89]
\draw [draw=none, fill=lightgray] (0,0) -- (1,1) -- (3,1) -- (4,0) -- (0,0);
\draw [draw=none, fill=lightgray] (0,8) -- (1,7) -- (3,7) -- (4,8) -- (0,8);
\draw [draw=none, fill=lightgray] (1,3) -- (0,4) -- (1,5) -- (3,5) -- (4,4) -- (3,3) -- (1,3);
\draw [thick, dashed] (1,5) -- (0,4) -- (1,3);
\draw [thick, dashed] (3,5) -- (4,4) -- (3,3);
\draw [thick, dashed] (1,1) -- (0,0);
\draw [thick, dashed] (3,1) -- (4,0);
\draw [thick, dashed] (1,7) -- (0,8);
\draw [thick, dashed] (3,7) -- (4,8);
\draw (1,1) -- (3,3); 
\draw (3,1) -- (1,3);
\draw (1,5) -- (3,7); 
\draw (3,5) -- (1,7);
\draw (2,0) -- (0,2) -- (4,6) -- (2,8) -- (0,6) -- (4,2) -- (2,0);
\node at (3,2) {\small I};	
\node at (2,2.5) {\small II};	
\node at (1,2) {\small III};	
\node at (2,1.5) {\small IV};	
\node at (2,3.5) {\small V};	
\node at (3,4) {\small VI};	
\node at (2,4.5) {\small VII};	
\node at (1,4) {\small VIII};	
\node [above right] at (3.4,2.4) {\small $J^+$};
\node [below right] at (3.4,1.6) {\small $J^-$};
\node [right] at (4,2) {\small $i^0$};
\end{tikzpicture} \qquad \qquad
\begin{tikzpicture}
\draw [draw=none, fill=lightgray] (2,0) -- (0,2) -- (2,4) -- (2,0);
\draw (2,0) -- (4,2) -- (2,4);
\draw [thick, dashed] (2,0) -- (0,2) -- (2,4);
\node at (2.7,2) {\small I};	
\node at (1.3,2) {\small II};	
\node [above right] at (3.4,2.4) {\small $J^+$};
\node [below right] at (3.4,1.6) {\small $J^-$};
\node [right] at (4,2) {\small $i^0$};
\end{tikzpicture}
	\caption{Conformal diagram of the spherically symmetric solution with the Lambda-function (\ref{alpha-family})
		and with the Schwarzschild radius $r_g (\infty) > r_c $ (left image) and $r_g (\infty) < r_c$ (right image),
		with $r_c$ given in (\ref{minr}).  Different regions are numbered in such a way that the 
		coordinate $t$ is timelike in the odd
		regions, and spacelike in the even ones. The flow of time is vertical in the white regions and 
		changes to horizontal in the grey regions. The boundaries between grey and white regions are 
		places where metric changes signature and, therefore, becomes singular. 
		Thick dashed lines indicate the true singularity, where $r = \infty$ and $\psi = \infty$.
		The configuration on the left image extends periodically and indefinitely upward and downward.
		We are living in one of the regions of type I; the asymptotic spatial infinity in this region is denoted
		by $i^0$, and the future and past null infinities are denoted by $J^+$ and $J^-$, respectively. \label{fig:bh}}
\end{figure}

As long as $\left| \Lambda_\psi \right| \ll 1$, the solution closely follows the Schwarzschild 
solution.  However, in the domain where $\left| \Lambda_\psi \right|$ becomes appreciable, 
it deviates from the Schwarzschild behavior.  It turns out to be more appropriate to describe 
solution in coordinates $(t, \psi)$ rather than $(t, r)$.  In this case, the solution can be extended
with coordinate $r$ being non-monotonic, and bounded from below.  For the theory defined by 
(\ref{alpha-family}), the corresponding conformal diagram is presented in Figure~\ref{fig:bh}, adopted
from \cite{Krasnov:2007ky}. 
Although the constructed metric (\ref{metric}) becomes singular on the boundaries between 
grey and white regions in the figure, the fundamental gauge field $A^i$ is regular there.  It is in this sense that 
we can speak about the `resolution' of Schwarzschild black-hole singularity in terms of fundamental connection 
variables $A^i$.  Note that the black-hole horizon in the metric exists only in the case \cite{Krasnov:2007ky}
\begin{equation}\label{minr}
r_g (\infty) > r_c \equiv \frac{ 2 \sqrt{\alpha}}{e \left(1 + \alpha \Lambda_0 / 3 \right)^{3/2}} \, .	
\end{equation}
In the opposite case, one obtains a solution resembling the Schwarzschild case with negative mass.

\section{Bianchi~I cosmology}

In general relativity, Bianchi~I cosmological model is described by a spatially Euclidean metric of the form
\begin{equation}
ds^2 = dt^2 - \left[ a_1 (t) dx^1 \right]^2 - \left[ a_2 (t) dx^2 \right]^2 -
\left[ a_3 (t) dx^3 \right]^2	\, ,
\end{equation}
where $a_i (t)$ are the corresponding scale factors.  The so-called Kasner regime is reached as $t \to 0$:
\begin{equation} \label{scale-gr}
a_i^2 \sim t^{2 p_i} \, , \qquad \sum_i p_i = 1 \, , \qquad \sum_i p_i^2 = 1 \, ,
\end{equation}
and describes approach to singularity at $t = 0$.  In this section, we will show how this singular behavior is `resolved'
in the modified theory with the cosmological function (\ref{alpha-family}).

By using the ${\rm SO} (3)$ gauge invariance of the theory, one can write the Bianchi~I ansatz for the connection
in the form
\begin{equation}
A^k = {\rm i} h_k (\tau) d x^k \quad \mbox{(no sum over $k$)}	\, ,
\end{equation}
where $\tau$ is, up to now, an arbitrary time parameter.  The
corresponding curvature two-form is
\begin{equation} \label{Fi}
F^i = d A^i + \frac12 \epsilon^{ijk} A^j \wedge A^k = {\rm i} \dot h_i d \tau \wedge d x^i -
\frac12 \epsilon^{ijk} h_j h_k d x^j \wedge d x^k \, ,
\end{equation}
where an overdot denotes derivative with respect to $\tau$. Calculating the wedge
product, we obtain
\begin{equation} \label{Xi}
F^i \wedge F^j = - 2 {\rm i} \delta^{ij} X_i h {\mathcal V}_c \, ,
\end{equation}
where ${\mathcal V}_c = d\tau \wedge dx^1 \wedge dx^2 \wedge dx^3 $ is the coordinate volume form,
$h =h_1 h_2 h_3$, and
\begin{equation} \label{eq-g}
X_i = \frac{\dot h_i}{h_i} \, .
\end{equation}
If we select the volume form ${\mathcal V}$ in (\ref{X}) to be
\begin{equation}
{\mathcal V} = - 2 {\rm i}  h {\mathcal V}_c \, ,
\end{equation}
then $X^{ij} = {\rm diag}\, \left(X_1, X_2, X_3 \right)$.  With this choice of the volume form, 
the pure-connection formulation equation (\ref{eqn-pure-conn}) reduces to the system
\begin{equation}\label{eq-diag}
\frac{d}{d \tau} \left( \frac{\partial {\mathcal L}}{\partial X_i}\right)  = {\mathcal L} (X) -  \frac{\partial
	{\mathcal L}}{\partial X_i} \sum_j X_j \, ,
\end{equation}
which is a system of first-order differential equations for $X_i$.  
By using time-reparametrisation freedom, it is always possible to choose the time variable
$\tau$ in such a way that ${\mathcal L} h = {\rm const}$.
With this choice, and using definition (\ref{eq-g}), 
one can integrate equation (\ref{eq-diag}) to obtain an implicit solution for $X (\tau)$:
\begin{equation} \label{Xsol}
\frac{\partial {\mathcal L} (X)}{\partial X_i} = {\mathcal L} (X) \left( \tau - \tau_i \right) \, ,
\end{equation}
where $\tau_i$ are arbitrary integration constants.  
Equations (\ref{Xsol}) and (\ref{eq-g}) give a complete solution to the problem for an
arbitrary theory from our class.
Without loss of generality, one can conveniently shift and normalize the time variable so that 
$\sum_i \tau_i = 0$  and $\sum_i \tau_i^2 = 6$.  

The canonical metric of this solution with volume form proportional to ${\mathcal L} (X) {\mathcal V}$ is 
given by
\begin{equation} \label{canonmet}
ds^2 =  \sqrt{ \left| \frac{{\mathcal L} (X) \prod_i X_i}{\Lambda_0 } \right| } \left[ - d \tau^2 +
\prod_j X_j^{-1}  \sum_k h_k^2 X_k \left(d x^k \right)^2 \right] \, .
\end{equation}

\subsection{Behavior in GR}

\begin{figure}[H]
	\centering
	\includegraphics[width=.7\textwidth]{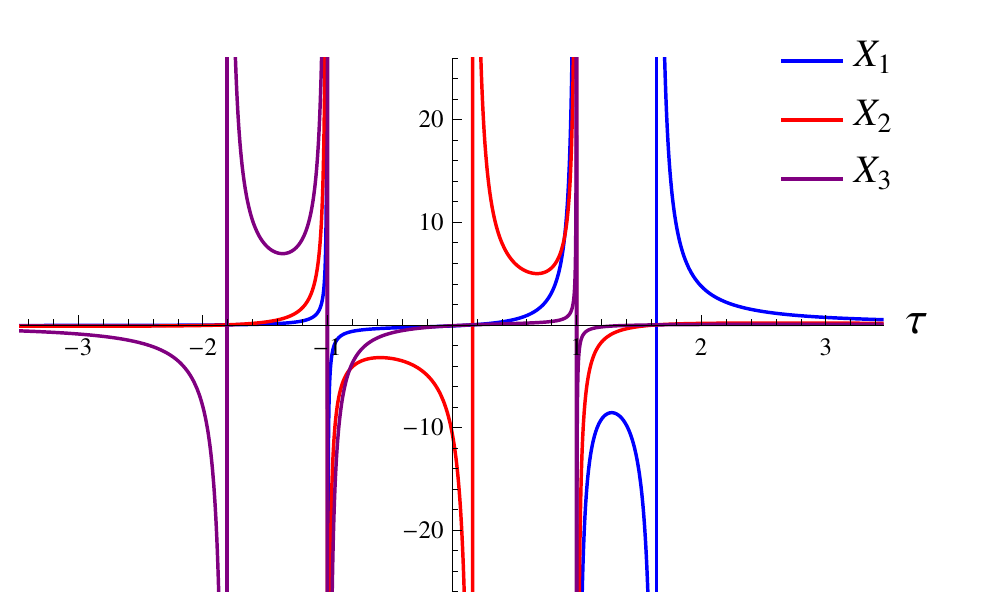}
	\caption{Behavior of the variables $X_i$ given by (\ref{Xsol}) in general relativity. Each variable
		$X_i$ has a pole at respective $\tau = \tau_i$. These are just special points of the solution, with all
		connection and metric components remaining finite. Furthermore, all variables $X_i$ have
		common poles at $\tau = \pm 1$. These are the singular points for both the connection and the
		metric. The region `between' two Kasner singularities describes a solution with negative
		$\Lambda$, while the two outer regions describe solutions with $\Lambda > 0$.} \label{fig:GR}
\end{figure}

The general solution for $X_i$ in the general relativity theory (\ref{GR}) is given by
\begin{equation}\label{X-GR}
X_1^{\rm GR} = \frac{(\tau-\tau_2)(\tau-\tau_3)}{\sqrt{3} (\tau^2-1)(\tau-\tau_1)}\, , 
\quad \mbox{etc.\@ with permutation of indices} \, .
\end{equation}
The quantities $X_i^{\rm GR}$ have
simple poles at $\tau = \tau_i$, and all blow up as $\tau \to \pm 1$, which
corresponds to the Kasner singularity. This behavior is illustrated in
Figure~\ref{fig:GR}.

Integrating (\ref{eq-g}) in the neighborhood of Kasner singularity and using (\ref{canonmet}), we 
see that the scale factors behave as
\begin{equation}
a_i^2 \sim (\tau - 1)^{p_i} \, ,
\end{equation}
where
\begin{equation}\label{ps}
p_1 = 1 - \frac{(1 - \tau_2)(1 - \tau_3)}{3 (\tau_1 - 1)} \, ,  \quad {\rm	etc} \, .
\end{equation}

If we arrange the time constants so that $\tau_3 < \tau_2 < \tau_1$, we obtain the following picture. 
As $\tau\to \infty$, we approach the De~Sitter
solution $X_i=1/{\sqrt{3} \tau}$. As time decreases, at $\tau=\tau_1$ we encounter a special
point where $X_1$ has a simple pole, while $X_2$ and $X_3$ vanish. Below this point, all
$X_i$ change sign, as does ${\mathcal L} (X)$. The component of the connection $h_1$ vanishes at
this point, while the components $h_2$ and $h_3$ remain finite. All components of the
canonical metric (\ref{canonmet}) remain finite at this point.  In terms of metric, this is the point
where the Hubble parameter $H_1 = \dot a_1 / a_1$ vanishes, and the scale factor $a_1$ reaches its 
minimum value.
As $\tau\to 1$ in our normalization, we approach the Kasner singularity, with the 
functions $X_i$ all
negative near the singularity, and all having a simple pole there. The scale factors $a_i^2$ exhibit the familiar
Kasner behavior (\ref{scale-gr}), with exponents $p_i$ given by (\ref{ps}). 

Since the gauge field diverges at the singularity, the domain $- 1 < \tau < 1$
in (\ref{X-GR}) can only be treated as another singular solution.  The solution in the time 
interval $\tau_2 < \tau < 1$ is described by GR with negative cosmological constant $- \Lambda_0$. 
The behavior near $\tau = 1$ is again Kasner.
The point $\tau_2$ again is a special point of the solution, in which $X_2$ has a simple
pole, while $X_1$ and $X_3$ have simple zeros. Thus, $h_2$ passes through zero at this
point, with $h_1$ and $h_3$ remaining finite and nonzero. The metric components are all
finite and non-zero. As $\tau\to -1$, we encounter another Kasner singularity. Thus,
this part of the solution interpolates between two Kasner singularities.
There is no asymptotic anti-De~Sitter regime in this case.

For $\tau < - 1$, we have another copy of asymptotically De~Sitter solution. We note
that all $X_i$ are positive near the singularity in this case, as is ${\mathcal L} (X)$. There is a
Kasner singularity as $\tau \to - 1$, and a special point at $\tau=\tau_3$ with
$h_3$ vanishing and all $X_i$ and ${\mathcal L} (X)$ changing sign. As $\tau\to -\infty$, we
approach another De~Sitter region. Since the time change $\tau \to - \tau$ makes the
region $\tau < - 1$ mathematically equivalent to the asymptotically De~Sitter region
$\tau > 1$ discussed above, it is clear that the Kasner exponents near the
singularity in the region $\tau < - 1$ are obtained from (\ref{ps}) by the
replacement $\tau_i \to - \tau_i$.

\subsection{Behavior in modified gravity}

In the modified theory of gravity described by the function $\Lambda (\Psi)$
in (\ref{pleb-modified}), it is quite difficult to calculate the corresponding Lagrangian ${\mathcal L} (X)$ 
in order to use the general solution (\ref{Xsol}).  In fact, this is difficult to do even in the 
simple case of modification (\ref{alpha-family}).  Therefore, to solve for the gauge-field dynamics, 
one has to resort to formalism with 
auxiliary matrix field $\Psi^{ij}$ together with the connection $A^i$ --- the formalism which is obtained
by excluding only the $B$-field from (\ref{pleb-modified}).  This procedure was followed by in our paper
\cite{Herfray:2015fpa}, where a general solution for the Bianchi~I case was obtained in the form
\begin{equation} \label{Xp}
X  = \frac{ T^{-2} \left(  {\rm Id} - \frac{\partial \Lambda (\Psi)}{\partial
		\Psi}\right)}{{\rm Tr\,} \left[ T^{-1}\left( {\rm Id} - \frac{\partial \Lambda (\Psi)}{\partial
		\Psi} \right) \right]} \, , \qquad 
\Psi = \Lambda (\Psi) \left[ \frac{T^{-1}}{{\rm Tr\,} (T^{-1})} -  \frac{1}{3}{\rm Id} \right]
\, .
\end{equation}
The last equation is to be solved with respect to $\Psi$, with the result to be substituted into
the first equation.  Here $T = {\rm diag} \left\{ (\tau - \tau_i) \right\}$, and ${\rm Id}$ is the identity matrix.

\begin{figure}[H]
	\centering
	\includegraphics[width=.7\textwidth]{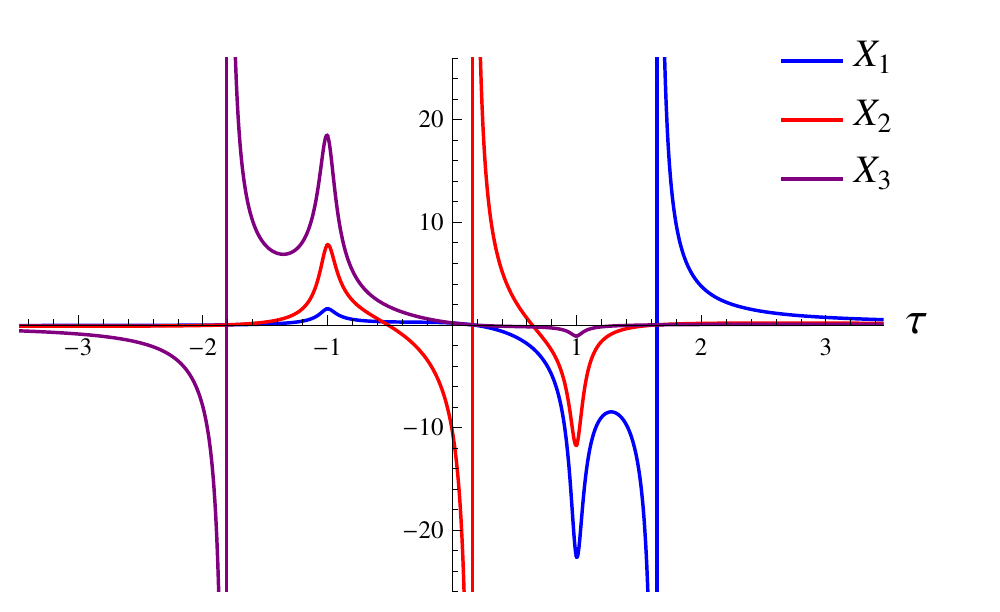}
	\caption{Behavior of the variables $X_i$ given by (\ref{Xp}) in the modified theory
		(\ref{alpha-family}) with $\alpha\Lambda_0 = 5 \times 10^{-3}$.  As in general relativity
		(see Figure~\ref{fig:GR}), each variable $X_i$ has a pole at respective $\tau = \tau_i$.
		However, the common poles at $\tau = \pm 1$ disappear, and the variables $X_i$
		become regular at these points.} \label{fig:modified}
\end{figure}

For the modified theory (\ref{alpha-family}), we observe that the singular behavior of the variables 
$X_i$ and connection-field components $h_i$ around $\tau = \pm 1$
in GR is replaced by a regular behavior of these quantities. 
The situation is depicted in Figure~\ref{fig:modified}.

In the general-relativistic solution, all $X_i$ are negative and blow up near the Kasner
singularity $\tau \to 1$. Our modification (\ref{alpha-family}) 
resolves this singularity, making all $X_i$ large negative but finite. Our solution thus
smoothly continues to the region  $\tau < 1$.  Since $X_i$ are finite and nonzero at
this point, ${\mathcal L}  (X)$ is also continuous.  However, the value of $\Lambda (\Psi)$ 
crosses zero and becomes negative at $\tau < 1$. 

As time decreases from $\tau = 1$, the next special point that we encounter is $\tau =
\tau_2$. 
We observe that $X_1$ and $X_3$ cross zero at $\tau = \tau_2$, while $X_2 \approx \sqrt{3}/(\tau
- \tau_2)$ has a simple pole, approaching positive infinity as $\tau\to \tau_2$ from
above. Since all $X_i$ were negative at $\tau = 1$, this means that $X_2$ crosses
zero at some $\tau_+ \in (\tau_2, 1)$. It then crosses zero once again at some
$\tau_- \in (- 1,\tau_2)$. This behavior is demonstrated in Figure~\ref{fig:modified}.
In the interval $(\tau_-, \tau_+)$, metric (\ref{canonmet}) changes signature from
$(-,+,+,+)$ to $(-,-,+,-)$, the spatial coordinate $x^2$ thus taking the role of time.
Thus, although we do not encounter singularity in the fundamental gauge field (all $h_i$
are everywhere smooth), there is a singularity in metric (\ref{canonmet}) at the points
$\tau = \tau_\pm$, where it also changes signature.

This behavior is in parallel to the `resolution' of the Schwarzschild singularity discussed 
in the preceding section, where we also observe change of signature of the corresponding metric.  
In both cases, it has to do with the fact that the value of $\Lambda (\Psi)$ becomes 
negative in our model (\ref{alpha-family}) as the quantity ${\rm Tr\,} \Psi^2$ increases.  Such a 
behavior of the `cosmological function' $\Lambda (\Psi)$ saves the theory from singularity.

\section{Discussion}

In this paper, we considered a specific family of chiral modified gravity theories in four dimensions. 
In the language of pure connection, the main idea is to consider general homogeneous functions 
of the curvature of the connection as Lagrangians of the theory.  Such a modification is 
guaranteed to lead to second-order field equations without new degrees of freedom.  The fundamental
field in this formulation is the connection field; the requirement of its curvature components to be (anti)-self-dual
then determines the conformal class of metrics.  The metric can thus be regarded as a derived, or secondary 
concept in this formulation.

The family of modifications under consideration is chiral; therefore, we are dealing with modifications 
of {\em complexified\/} GR\@. Such a theory begs for reality conditions, which are not difficult to formulate 
in the cases of Riemannian and split signatures of the arising metric.  In the case of physical Lorentzian 
signature, no general reality conditions are known.  Another closely related problem is coupling of this 
theory to matter. Note that matter fields in our setting should couple to the fundamental field describing 
gravity, which is the connection field.  Alternatively, one can start with a more general family of gauge 
theories \cite{Krasnov:2011hi} that describe gravity as well as matter fields.  The problem of reality 
conditions will remain in any setting; since matter fields couple to a complex-valued connection, some 
reality conditions are required to make sense of the arising dynamics.

In some special cases, such as the spherically symmetric case or the Bianchi~I setup, the problem of 
reality causes no difficulty: the self-dual part of the Weyl curvature is automatically real, resulting in 
real effective metrics.  Our main finding in these cases is that a natural one-parameter family of modifications
(\ref{alpha-family}) with positive parameter $\alpha$ resolves the Schwarzschild black-hole and Kasner 
singularities. In the spherically symmetric case, we obtain a space-time extending indefinitely and periodically 
beyond the would-be black-hole singularity (see Figure~\ref{fig:bh}).  In the case of Bianchi~I symmetry, 
we obtain a solution bypassing the would-be Kasner singularities and connecting two 
asymptotically De Sitter regions (see Figure~\ref{fig:modified}).  In both space-times, the fundamental 
connection field remains regular, although the related metric has singularity points with changes
of signature along some coordinates.  This type of metric singularity, in
the absence of singularity in the basic connection field, appears to be generic to
the family of modified theories under investigation.
The fact that chiral modified gravity theories can resolve the singularities of general-relativistic solutions 
is quite remarkable and makes them worth further investigation.  

%

\vspace{6pt} 


\acknowledgments{The work of Yu.~S. was supported in part by grant 6F of the Department of Target 
Training of the Taras Shevchenko Kiev National University under the National Academy of Sciences of Ukraine.}


\conflictsofinterest{The authors declare no conflict of interest.
The founding sponsors had no role in the design of the study; in the collection, analyses, 
or interpretation of data; in the writing of the manuscript, and in the decision to publish the results.} 

\abbreviations{The following abbreviations are used in this manuscript:\\

\noindent 
\begin{tabular}{@{}ll}
GR & General Relativity \\
DOF & Degrees of Freedom
\end{tabular}}

\end{document}